\documentclass[a4paper]{jpconf}
\usepackage{graphicx}
\newcommand{\be}{\begin{equation}}
\newcommand{\ba}{\begin{eqnarray}}
\newcommand{\ee}{\end{equation}}
\newcommand{\ea}{\end{eqnarray}}
\newcommand{\pa}{\partial}

\begin{document}
\title{$D$-meson diffusion in hadronic matter}

\author{Juan M Torres-Rincon$^{1,}$\footnote[0]{Speaker}, Luciano M Abreu$^2$, Daniel Cabrera$^3$, \\ Felipe J Llanes-Estrada$^4$, Laura Tolos$^{1,3}$}
\address{$^1$ Institut de Ci\`encies de l'Espai (IEEC/CSIC), Campus Universitat Aut\`onoma
de Barcelona, Facultat de Ci\`encies, Torre C5, E-08193 Bellaterra, Spain}
\address{$^2$ Instituto de F\'{\i}sica, Universidade Federal da Bahia, 40210-340,
Salvador, BA, Brazil}
\address{$^3$ Frankfurt Institute for Advances Studies. Johann Wolfgang Goethe University, Ruth-Moufang-Strasse 1,
60438 Frankfurt am Main, Germany}
\address{$^4$ Departamento de F\'{\i}sica Te\'orica I, Universidad Complutense, 28040
Madrid, Spain}

\begin{abstract}
We present effective-field-theory results with unitarized interactions on the $D$-meson transport coefficients in a gas populated
by light mesons and baryons at finite temperature and baryochemical potential. The Fokker-Planck equation is used
to compute the drag force, the relaxation time and the diffusion coefficients of $D$ mesons for collisions at FAIR.
At finite baryochemical potential, the combined effect of net baryonic density and sizable meson-baryon interaction makes the $D$ mesons
to relax more efficiently than in the $\mu_B=0$ case. We also describe the connection with the quark-gluon plasma phase in adiabatic
trajectories on the phase diagram at both zero and finite baryochemical potential.
\end{abstract}

\section{Introduction}

   Some of the most useful tools to study the quark-gluon plasma (QGP) phase --formed in the early stages of a relativistic heavy-ion collision--
are heavy quarks. Due to their large mass ($m_c,m_b \gg \Lambda_{QCD},T$) they can be treated as Brownian particles interacting with lighter partons.
Not only the average momentum of the heavy quarks is modified by the collisions, but also the momentum
distribution is broadened. The efficiency of these two processes is quantified by the {\it drag force} and the {\it momentum-space diffusion
coefficient}, respectively. 

   The previous picture also holds after hadronization. The role of the Brownian particle is played by a $D$ or $B$ meson,
whereas the thermal bath is now composed of pions, kaons, nucleons and other light hadrons with whom the heavy meson may interact.
Thus, the transport coefficients of $D$ mesons~\cite{Abreu:2011ic,Tolos:2013kva} and $B$ mesons~\cite{Abreu:2012et} become of key importance in the
equilibration of these hadrons. The presence of these coefficients will amount on the modification of the final heavy meson
spectra and affect the flow harmonics $v_n$ and the nuclear suppression factor $R_{AA}$. To study these transport coefficients and quantify the energy loss of heavy quarks in the hot and dense medium
we use quantum kinetic theory.

   We focus on the heavy-ion physics to be produced at the Facility for Antiproton and Ion Research (FAIR)~\cite{Friman:2011zz}, where
the collision energy will be lower than that for collisions at the Large Hadron Collider, for instance. Lower energies typically produce a net baryon density
in the central rapidity region. This fact will imply physics at non-zero baryochemical potential in the QCD phase diagram,
maybe close to the conjectured critical point. For these collisions, the available energy density makes difficult the production
of $B$ mesons. Therefore, in this manuscript we focus our attention to the calculation of $D$-meson transport coefficients~\cite{Tolos:2013kva}.

In Sec.~\ref{sec:fp} we introduce the Fokker-Planck kinetic equation and detail the physical interpretation of the $D$-meson
transport coefficients. In Sec.~\ref{sec:int} we briefly describe the interaction of $D$ mesons with lighter hadrons, both mesons and baryons.
Finally, in Sec.~\ref{sec:result} we present our main results and conclusions.

\section{Fokker-Planck equation and transport coefficients\label{sec:fp}}

  The collective description of $D$ mesons is encoded in their distribution function $f(t,p)$. The temporal evolution of $f(t,p)$ follows a kinetic
equation defined in the phase space, which in the case of a Brownian particle in a thermal bath composed of lighter particles is the {\it Fokker-Planck equation}:
\be \label{eq:fp} \frac{\pa f(t,{\bf p})}{\pa t} = \frac{\pa}{\pa p_i} \left\{ F_i (\mathbf{p}) f(t,{\bf p}) + \frac{\pa}{\pa p_j} \left[ \Gamma_{ij} (\mathbf{p}) f(t,\bf{p}) \right] \right\} \ ,\ee
where $i=1,2,3$ is the spatial index, $F_i$ is the drag force or friction term accounting for the average momentum change, and $\Gamma_{ij}$
is the momentum-space diffusion matrix  which accounts for the broadening of the momentum distribution. The computation of these coefficients is performed through their microscopic expressions
in terms of momentum averages, \be F_i (\mathbf{p})= \int d\mathbf{k} \ w(\mathbf{p},\mathbf{k}) \ k_i \ , \qquad \Gamma_{ij} (\mathbf{p})= \frac{1}{2} \int d\mathbf{k} \ w(\mathbf{p},\mathbf{k}) \ k_i k_j \ , \ee
where the interaction measure $w(\mathbf{p},\mathbf{k})$ represents the probability of a $D$ meson with momentum $\mathbf{p}$ to suffer a collision
with a light particle, loosing a momentum $\mathbf{k}$:
\ba w(\mathbf{p},\mathbf{k})  &=&  g_l \int \frac{d^3q}{(2\pi)^9} \ n_{F,B} (E_l(q),T) \ \left[ 1\pm n_{F,B} (E_l(q+k),T) \right]  \frac{1}{2E_D(p)} \frac{1}{2E_l(q)} 
\frac{1}{2E_D(p-k)}  \nonumber \\ 
& \times & \frac{1}{2E_l(q+k)} (2\pi)^4 \delta (E_D(p)+E_l(q)-E_D(p-k)-E_l(q+k)) \ \overline{|\mathcal{M}|^2} \ , 
\label{eq:omega}
\ea
with $g_l$ being the spin-isospin degeneration of the light particle, $n_{F,B}$ its equilibrium distribution function following Fermi-Dirac or Bose-Einstein statistics, respectively; $E_D(p)$ is the energy of the $D$ meson with momentum $\mathbf{p}$,
$E_l(q)$ the energy of the light particle with momentum $\mathbf{q}$ and $\overline{|\mathcal{M}^2|}$ is the average scattering matrix element squared of the binary collision.

If we consider an isotropic gas and focus on the {\it static limit} (where $p\rightarrow 0$) there is only one independent transport coefficient. For definiteness,
we take it to be the drag force $F(p)=F_i p^i/p^2$, while the momentum-space diffusion coefficient $\Gamma$ is related to $F$ through the Einstein relation $\Gamma=Fm_DT$ (where $m_D$ is the $D$-meson mass and $T$ the temperature of the thermal medium).

In several talks at this workshop some emphasis was put on the concept of {\it relaxation time}. This coefficient describes how efficient the relaxation 
of the non-equilibrium distribution function is, while it relaxes towards the equilibrium state (Boltzmann's H theorem). Consider for
simplicity the one dimensional case of Eq.~(\ref{eq:fp}) and assume constant transport coefficients $F$ and $\Gamma$. The analytical solution of the kinetic equation
is
\be f(t,p) = \sqrt{\frac{F}{2\pi \Gamma (1-e^{-2Ft})}} \exp \left[ - \frac{F}{2 \Gamma} \frac{(p-p_0 e^{-Ft})^2}{1-e^{-2Ft}}   \right] \ , \ee
where $p_0$ is the initial $D$-meson momentum. It is immediate to see that $F^{-1}$ should play the role of an intrinsic time for the relaxation process.
In particular, the average momentum reads
\be \langle p \rangle \equiv  \int_{-\infty}^{\infty} dp \ p f(t,p) \ \Big/  \int_{-\infty}^{\infty} dp \ f(t,p) = p_0 \ e^{-\frac{t}{\tau_R}} \ , \ee
where we have defined the relaxation time $\tau_R \equiv 1/F$, which provides a characteristic time accounting for the exponential decay of the average momentum.

  We can also introduce the {\it diffusion coefficient in position space $D_x$}. The standard result for the average square displacement
\be \langle (x(t) - x_0)^2 \rangle = 2D_x t \ , \ee
shows that $D_x$ is a measure of the ``speed'' of the diffusion in space. A large $D_x$ implies that the $D$ meson can diffusively travel larger
distances for a given time. In the static limit this coefficient is related to the drag force by the relation $D_x = T / (m_D F)$.

All these coefficients depend on the temperature and baryochemical potential. In a heavy-ion collision both variables change with time along
the fireball's trajectory in the phase diagram. For simplicity, we assume adiabatic trajectories --with a constant entropy per baryon consistent
with the energies expected at FAIR collisions-- which give us the implicit relation $\mu_B=\mu_B(T)$.

\section{Interaction and effective field theories\label{sec:int}}

   The interaction between the $D$ mesons and the light hadrons --as it lies in the nonperturbative and confined regime of QCD-- can be
described by effective models. The global symmetries of the QCD Lagrangian should be implemented into the effective models 
allowing a reliable description of the hadronic interaction. We take the {\it chiral and heavy-quark-spin symmetries} to be the fundamental
guides in order to construct our effective Lagrangian (therefore, it results in a double expansion). 
In addition to these symmetries, our third principle is the exact satisfaction of the {\it unitarity condition} of the scattering matrix.

   To account for the $D$ meson--light meson scattering we introduce pions, kaons, antikaons and $\eta$ mesons. Details on the effective
theory based on chiral and heavy-quark-spin expansions can be found in Refs.~\cite{Abreu:2011ic,Tolos:2013kva,Guo:2009ct}. Using Feynman rules
we construct the perturbative amplitude (or potential) $V$ for each scattering channel.
In the meson-baryon sector we also construct the effective Lagrangian based on heavy-quark-spin symmetry which accounts for the interaction of $D$ mesons with
nucleons and $\Delta$ baryons. Details on the effective theory and the computation of the potential $V$ can be found in~\cite{Tolos:2013kva,Mizutani:2006vq,GarciaRecio:2008dp,Garcia-Recio:2013gaa,Romanets:2012hm}.

In both sectors we apply a unitarization method~\cite{Oller:1997ti} (based on a Bethe-Salpeter (BS) equation) to extend the interaction description up to temperatures $T\simeq 140$ MeV without loosing physical sense.
In addition, we are able to dynamically generate resonant states like the $D_0 (2400)$ in the meson--meson sector and the $\Sigma_c (2800)$ and $\Lambda_c (2595)$ in 
the meson--baryon sector, which are automatically taken into account in the interaction. The scattering matrix element in Eq.~(\ref{eq:omega}) is related to the perturbative amplitude as:
\be \mathcal{M} (s) \propto \frac{V(s)}{1-G(s)V(s)} \ , \ee
where $G$ represent the diagonal matrix of the two intermediate propagators in the BS equation.

\section{Results and Conclusions\label{sec:result}}

\begin{figure}[h]
\includegraphics[width=0.46\textwidth]{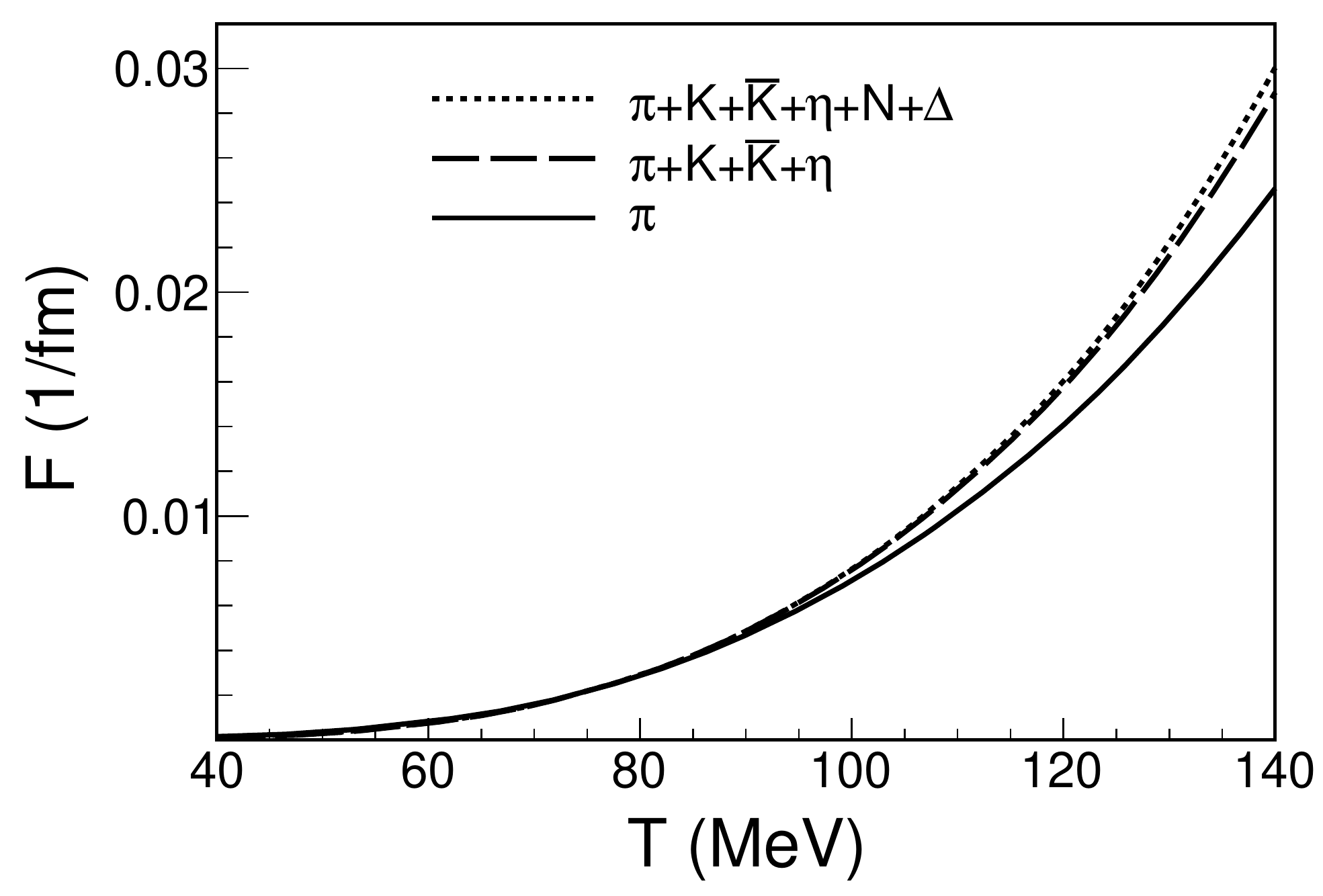}\hspace{2pc}%
\begin{minipage}[b]{18pc}
\caption{\label{fig:coeff} Drag force $F(p)$ for a $D$ meson with momentum $p=100$ MeV as a function of temperature for several species in the
thermal bath at $\mu_B=0$ (LHC collisions). Clearly the dominant contribution comes from pions as they are the most abundant in the bath. The contribution of baryons is practically negligible.
The diffusion coefficient in momentum space is related to this coefficient via the Einstein relation $\Gamma=Fm_D T$. \vspace{0.3cm}}
\end{minipage}
\end{figure}

   We present our main results taken from Ref.~\cite{Tolos:2013kva}. In Fig.~\ref{fig:coeff} we show the drag coefficient $F(p=100 \textrm{ MeV})$ for a $D$ meson as a function of
the temperature when the baryochemical potential is set to zero. The largest contribution comes from the pions, as these mesons populate the gas at 
temperatures below the hadronization one. The kaons, antikaons and $\eta$ mesons contribute appreciably only at high temperatures
$T>100$ MeV, being the baryon contribution ($N+\Delta$) negligible at $\mu_B=0$.

\begin{figure}[h]
\includegraphics[width=0.46\textwidth]{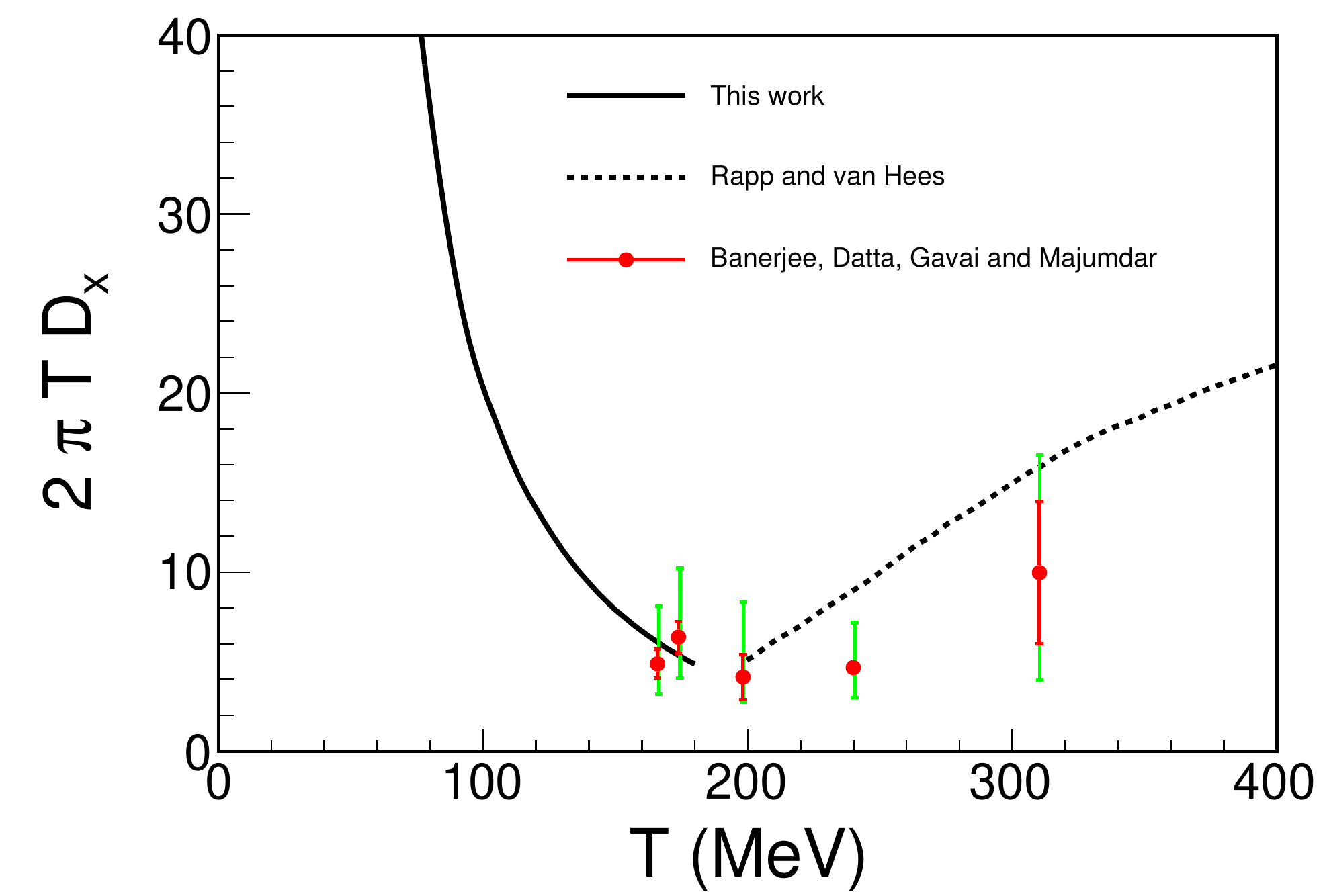}\hspace{2pc}%
\begin{minipage}[b]{18pc}\caption{\label{fig:dxmu0} Diffusion coefficient multiplied by the thermal wavelength $2\pi T$ as a function of temperature,
around the crossover temperature at $\mu_B=0$. The solid line represents our results for hadronic matter with all species included. The dashed 
line shows the result of Ref.~\cite{Rapp:2009my} for a quark-gluon plasma. Dots represent the result from the lattice-QCD calculation extracted from Ref.~\cite{Banerjee:2011ra}.} \vspace{0.7cm}
\end{minipage}
\end{figure}

  The position-space diffusion coefficient $D_x$ is shown at $\mu_B=0$ in Fig.~\ref{fig:dxmu0}, when all hadrons are included in the calculation.
We normalize this coefficient by the thermal wavelength $2\pi T$ to construct an adimensional number. The calculation for the QGP (Rapp and Hees) is
a $T$-matrix calculation from Ref.~\cite{Rapp:2009my} and the lattice-QCD computation (Banerjee, Datta, Gavai and Majumdar) is taken from Ref.~\cite{Banerjee:2011ra}.

\begin{figure}[h]\hspace{2pc}%
\includegraphics[width=0.38\textwidth,height=5cm]{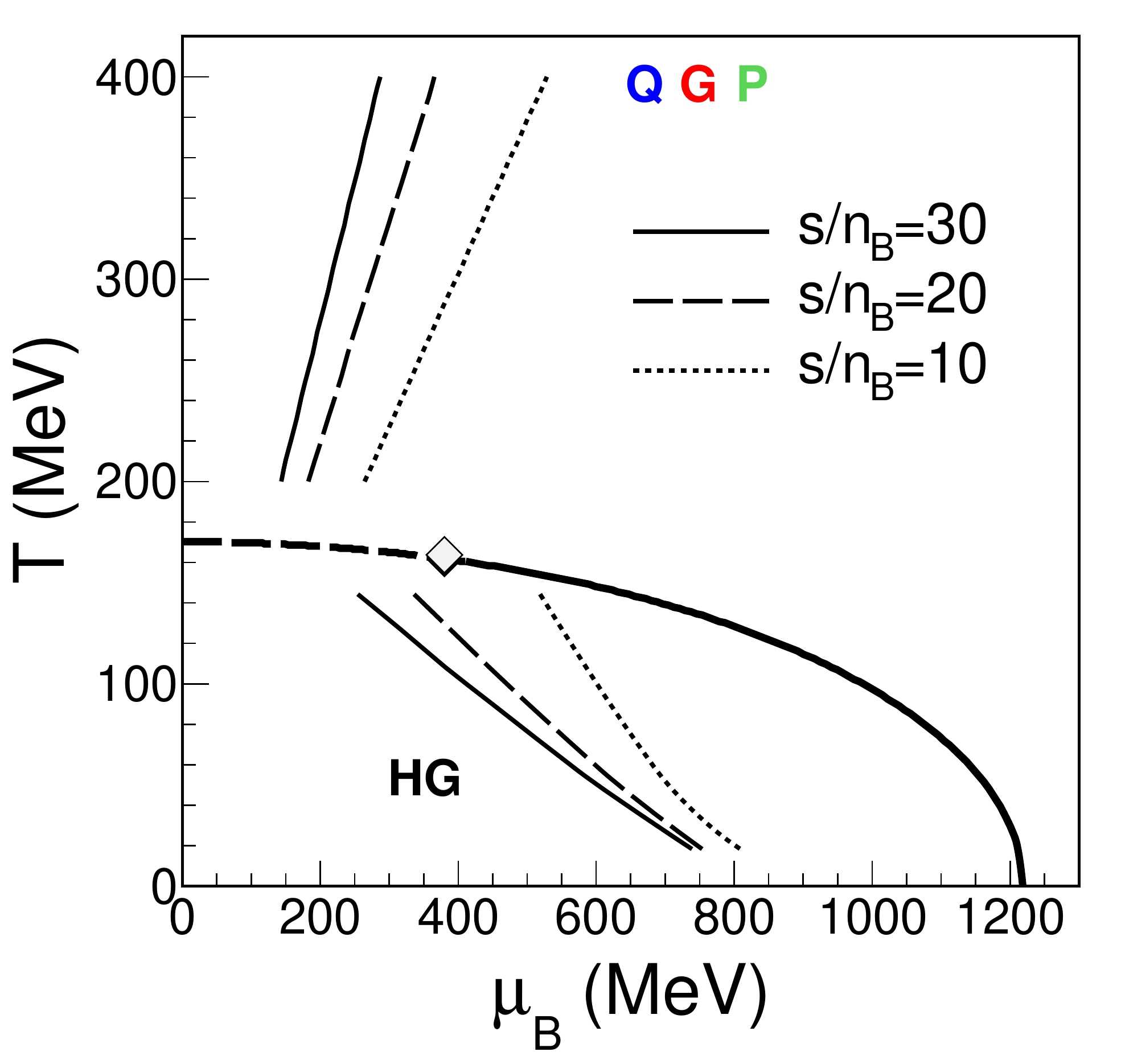}\hspace{2pc}%
\includegraphics[width=0.45\textwidth]{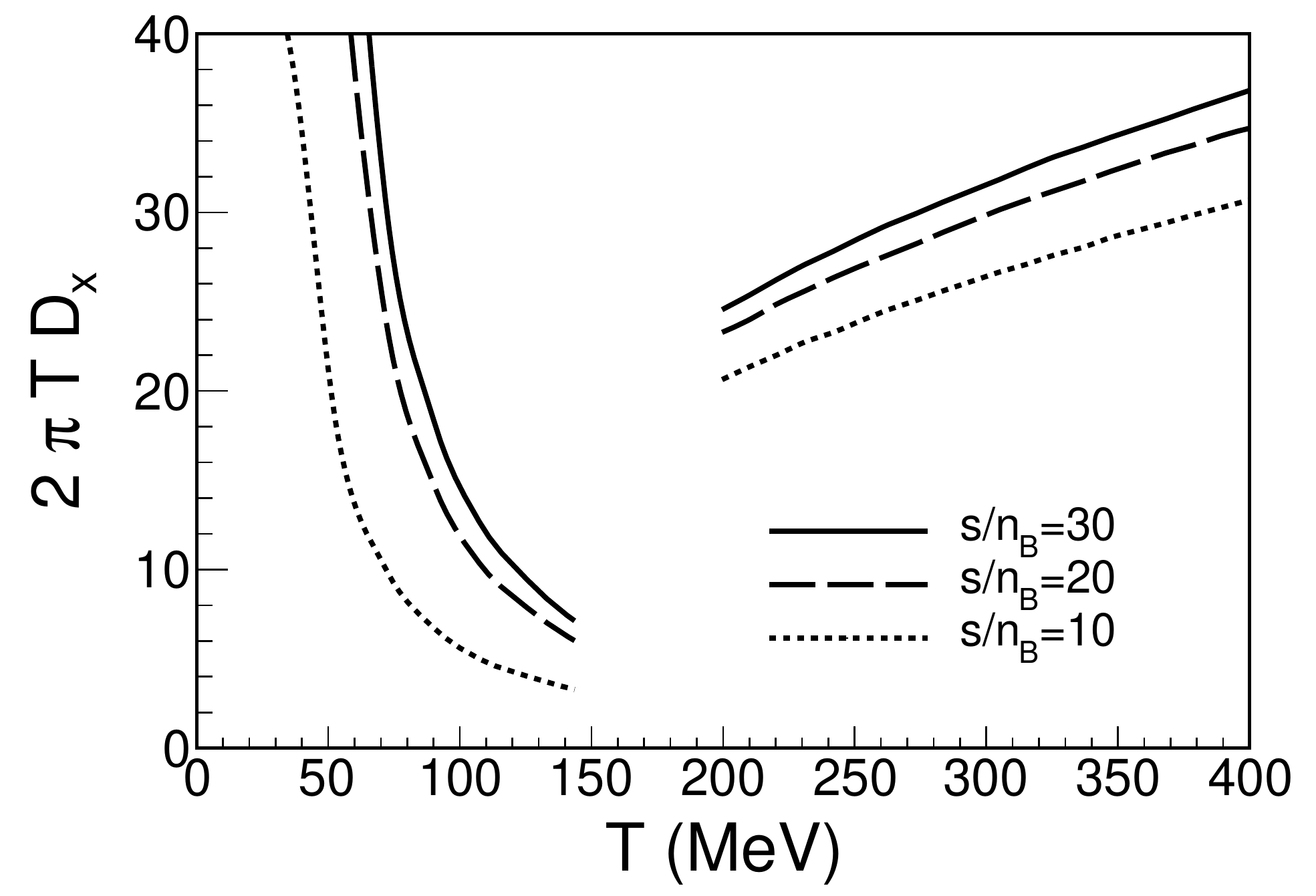}
\caption{\label{fig:traj} Left panel: Adiabatic trajectories for different initial conditions at FAIR collisions.
Right panel: $D_x$ as a function of temperature for the three adiabatic trajectories.}
\end{figure}

Turning to the $\mu_B\neq 0$ case --relevant for the physics at FAIR-- we consider three characteristic adiabatic trajectories in the 
phase diagram for different initial collision energies ($\sqrt{s} \approx 5-40A$ GeV~\cite{Friman:2011zz}). In the left panel of Fig.~\ref{fig:traj} we sketch our
trajectories on the phase diagram. They correspond to fixed entropy per baryon of $10-30$~\cite{Bravina:2008ra}. In the right panel we plot our results in the 
hadron gas (at low temperature) and the results of a perturbative QGP at finite quark chemical potential based on the computation
of Ref.~\cite{Moore:2004tg}.

\begin{figure}[h]
\includegraphics[width=0.44\textwidth]{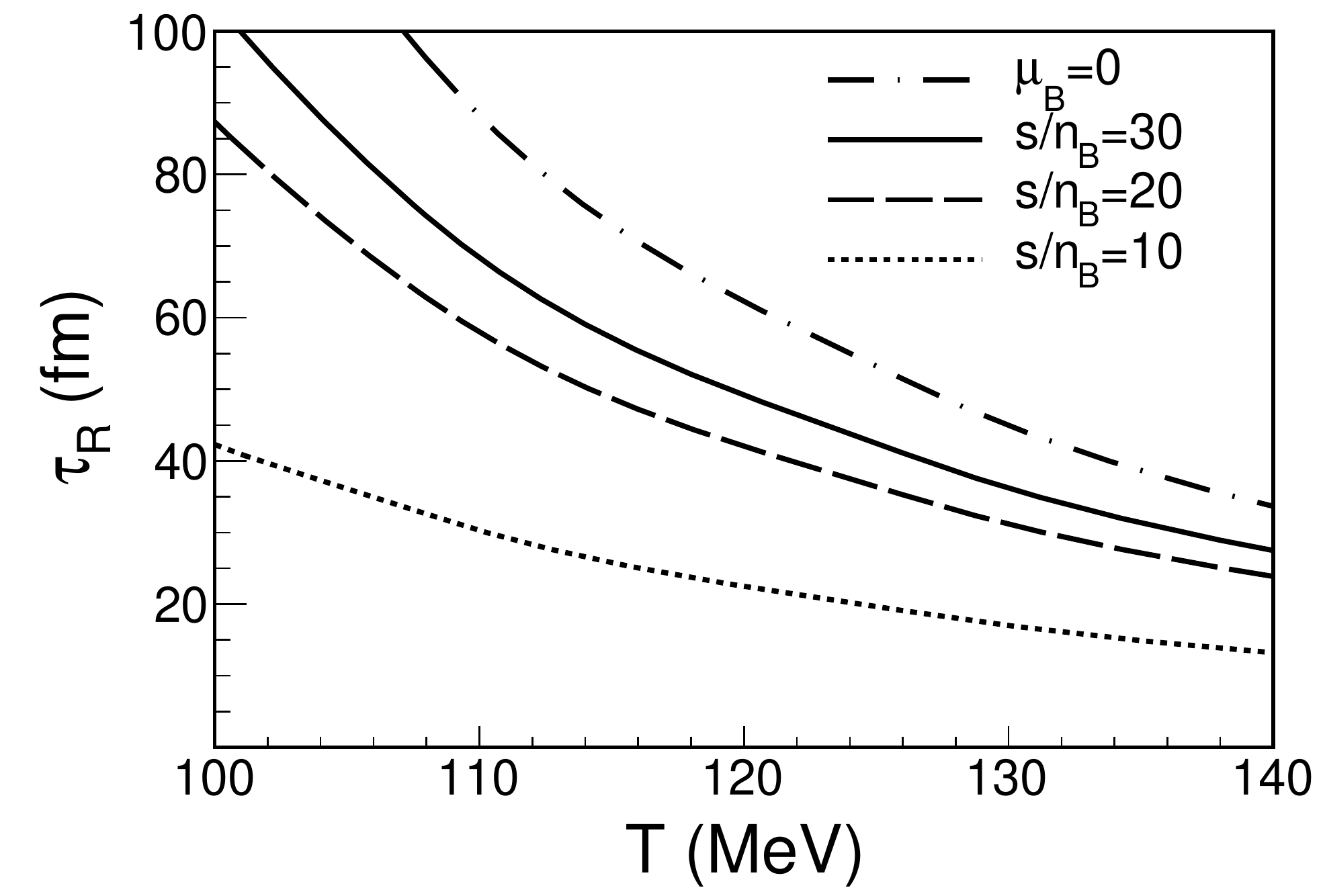}\hspace{2pc}%
\begin{minipage}[b]{18pc}\caption{\label{fig:rel} Relaxation time for the adiabatic trajectories as a function of temperature. 
The trajectories with larger baryonic density (those with smaller entropy per baryon) make the $D$ meson to have a lower relaxation time and hence,
to follow a more efficient relaxation via baryonic collisions. \vspace{1.2cm}}
\end{minipage}
\end{figure}

 Finally, we show the relaxation time for each adiabatic trajectory plus the result at $\mu_B=0$ as a function of temperature in Fig.~\ref{fig:rel}.
Notice that the presence of baryons at finite $\mu_B$ enhances the $D$-meson stopping and the relaxation time is accordingly reduced.
Therefore collisions at FAIR put the $D$ mesons closer to the equilibrium state than at the LHC.
These relaxation times should be compared to the fireball's lifetime which is of the order of $t_f \sim 10$ fm. In any case, the $D$ mesons 
cannot totally relax through collision as $\tau_R > t_f$, confirming that $D$ mesons should be sensible to some properties of the initial QGP.

  In summary, we have presented our latest results on the transport coefficients of $D$ mesons that describe the 
energy loss and diffusion of these states in a hadronic medium at finite temperature and baryochemical potential. Focusing
on FAIR physics, we have obtained that the collisions with baryons make the $D$ mesons more thermally relaxed than the case at 
vanishing baryochemical potential. We have obtained a nice connection with the results of QGP at high temperature, providing an
indication for a possible minimum of $2\pi TD_x$ at both the crossover temperature (at $\mu_B=0$) and at the first-order transition temperature
(at FAIR collisions). 

\ack JMTR acknowledges the organizers of FAIRNESS 2013 for the hospitality received and the opportunity to
present these results. This work has been financed by grants FPA2010-16963, FPA2011-27853-C02-01, FPA2011-27853-C02-02 (Spain), BMBF (Germany) under
project no. 05P12RFFCQ, EU Integrated Infrastructure Initiative Hadron Physics Project under Grant Agreement 
n.~227431 and Grant No. FP7-PEOPLE-2011-CIG under Contract No. PCIG09-GA-2011-291679.
LMA thanks CAPES (Brazil) for partial financial support. DC acknowledges support from Centro Nacional de Fisica de Part\'iculas,
Astropart\'iculas y Nuclear (Consolider - Ingenio 2010) and LT acknowledges support from the Ram\'on y Cajal Research Programme.

\end{document}